\documentclass[conference]{IEEEtran}
\IEEEoverridecommandlockouts
\usepackage{cite}
\usepackage{amsmath,amssymb,amsfonts}
\usepackage{algorithmic}
\usepackage{graphicx}
\usepackage{textcomp}
\usepackage{xcolor}
\usepackage{url} 
\usepackage{subcaption} 
\usepackage{hyperref} 
\hypersetup{
    hidelinks=true 
}

\def\BibTeX{{\rm B\kern-.05em{\sc i\kern-.025em b}\kern-.08em
    T\kern-.1667em\lower.7ex\hbox{E}\kern-.125emX}}
\begin{document}

\title{
Explainable AI model reveals disease-related mechanisms in single-cell RNA-seq data
}

\author{
\IEEEauthorblockN{
Mohammad Usman\textsuperscript{1,2,3,4},
Olga Varea\textsuperscript{2,3,4},
Petia I Radeva\textsuperscript{1},
Josep M Canals\textsuperscript{2,3,4},
Jordi Abante\textsuperscript{1,2,3,4},
Daniel Ortiz\textsuperscript{1}\IEEEauthorrefmark{1}
}

\IEEEauthorblockA{\textsuperscript{1}
\textit{Dept. of Mathematics \& Computer Science}, 
Universitat de Barcelona, Barcelona, Spain}

\IEEEauthorblockA{\textsuperscript{2}
\textit{Dept. of Biomedical Sciences}, 
Faculty of Medicine and Health Sciences, 
Institute of Neuroscience, \\ 
University of Barcelona, Barcelona, Spain}

\IEEEauthorblockA{\textsuperscript{3}
\textit{Creatio, Production and Validation Center of advanced therapies}, 
Universitat de Barcelona, Barcelona, Spain}

\IEEEauthorblockA{\textsuperscript{4}
\textit{Institut d’Investigacions Biomèdiques August Pi i Sunyer (IDIBAPS), Barcelona, Spain}
}
\IEEEauthorblockA{\IEEEauthorrefmark{1}
Corresponding author: daniel.ortiz@ub.edu}
}

\maketitle

\begin{abstract}
Neurodegenerative diseases (NDDs) are complex and lack effective treatment due to their poorly understood mechanism. The increasingly used data analysis from Single nucleus RNA Sequencing (snRNA-seq) allows to explore transcriptomic events at a single cell level, yet face challenges in interpreting the mechanisms underlying a disease. On the other hand, Neural Network (NN) models can handle complex data to offer insights but can be seen as black boxes with poor interpretability. In this context, explainable AI (XAI) emerges as a solution that could help to understand disease-associated mechanisms when combined with efficient N models. However, limited research explores XAI in single-cell data. In this work, we implement a method for identifying disease-related genes and the mechanistic explanation of disease progression based on  model combined with SHAP. We analyze available Huntington's disease (HD) data to identify both HD-altered genes and mechanisms by adding Gene Set Enrichment Analysis (GSEA) comparing two methods, differential gene expression analysis (DGE) and NN combined with SHAP approach. Our results show that DGE and SHAP approaches offer both common and differential sets of altered genes and pathways, reinforcing the usefulness of XAI methods for a broader perspective of disease.  
\end{abstract}


\section{Introduction}

In recent years, transcriptome analysis has emerged as a vital tool for unraveling how gene expression contributes to physiological and pathological conditions \cite{stubbington2017single}. Leveraging new high-throughput RNA sequencing techniques, such as single-cell or single-nuclei RNA sequencing (sc/snRNA-seq), researchers can examine gene expression patterns at the granularity of individual cells. Traditional differential gene expression (DGE) analysis methods often face challenges in identifying the complex relationships between genes and pathological phenotypes \cite{ij2018statistics, das2021comprehensive} due to limitations to handle the multi-dimensional nature of transcriptomic data and thus, to capture the gene associations and patterns that contribute to the onset and progression of a disease. Traditional approaches are usually based on generalized linear models (GLMs), using Poisson or negative binomial families, which might face difficulties when dealing with single-cell data characterized by high sparsity and dropout events, potentially leading to over or underestimation of gene expression differences. More sophisticated models, such as zero-inflated or mixture models, attempt to address these issues. 

Machine learning (ML) approaches constitute an alternative to standard DGE methods due to their ability to learn complex patterns from high-dimensional data. Neural Network (NN) is a class of machine learning models inspired by the function of the human brain. An NN consists of multiple processing neural layers with a very large number of learnable parameters that help approximate non-linear transformations able to learn informative data representation at different scales. Unlike traditional Machine Learning (ML) models, NNs can automatically learn features from input data with end-to-end learning mechanisms \cite{lecun2015deep}. However, despite NN models effectiveness, they are often considered a "black-box": it is challenging to interpret their predictions and understand the learning process. This lack of transparency is a significant barrier to gaining insights from the ML process. To address this issue, explainable artificial intelligence (XAI) techniques are used to understand ML mechanisms. Thus, XAI play an important role in uncovering the underlying mechanisms driving disease progression learned by the NN. Here, we present the application of both NNs and XAI to identify altered disease-associated genes that underlie the possible mechanisms of a given disease of interest and compare the results obtained with differential gene expression as a traditional method.

\begin{figure*}[t]
    \centering
    \begin{subfigure}[b]{0.45\textwidth}
        \centering
        \includegraphics[width=\textwidth]{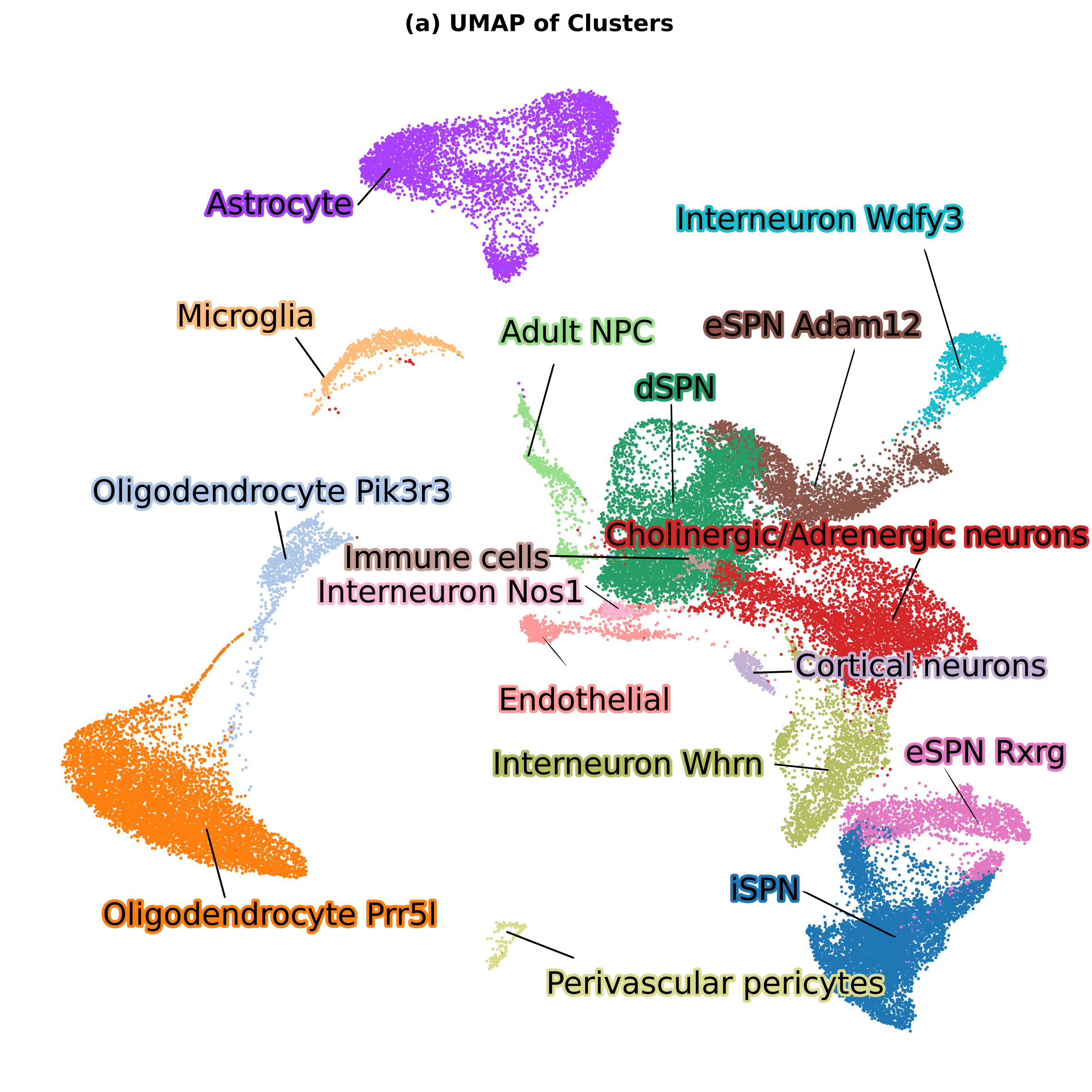}
    \end{subfigure}
    \hfill
    \begin{subfigure}[b]{0.48\textwidth}
        \centering
        \includegraphics[width=\textwidth]{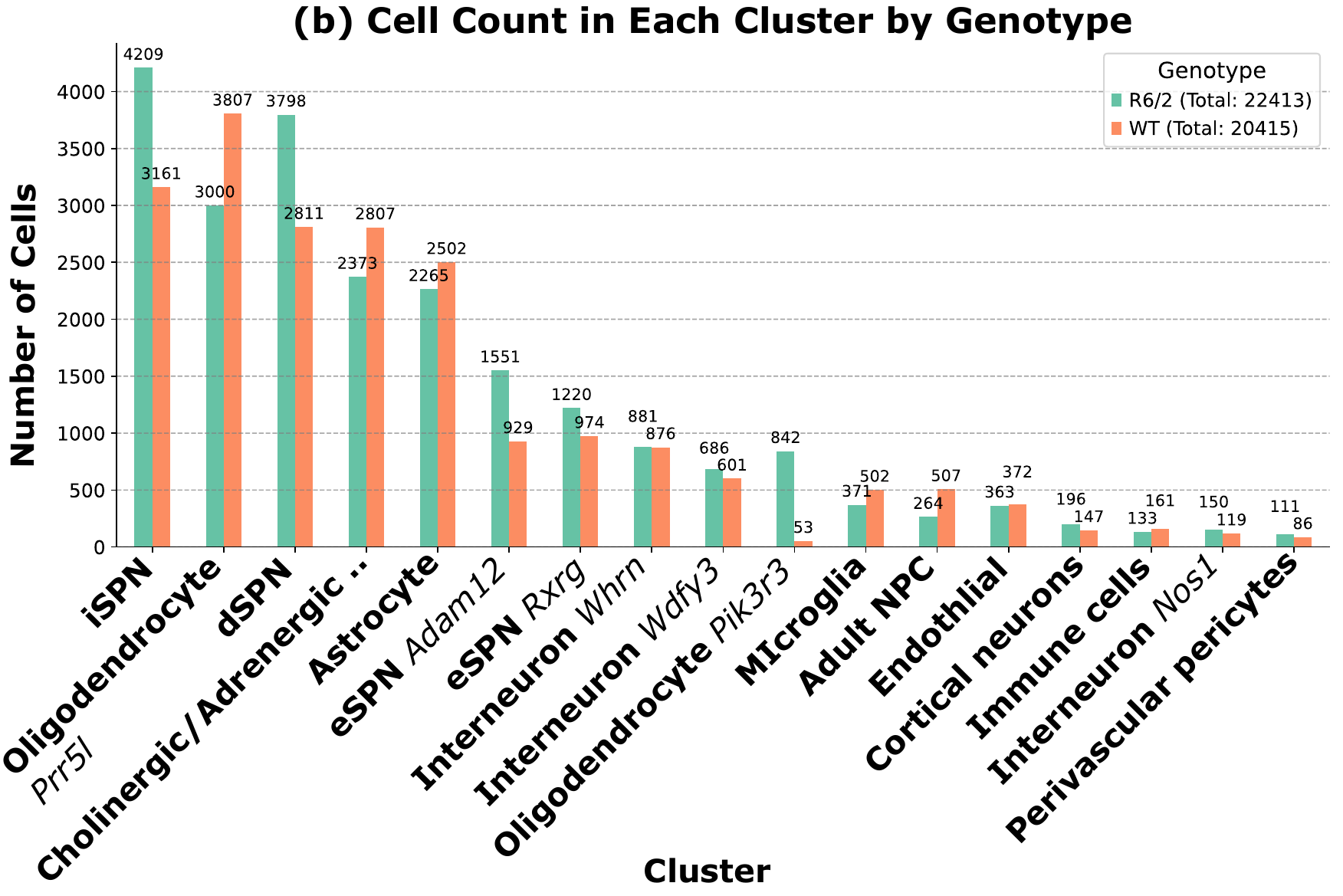}
    \end{subfigure}
    \caption{\textbf{Single-nuclei RNA-seq mouse data}. \textbf{a.} Integration of single-nuclei RNA-seq data with colors representing the cell-type identified using cluster markers. Here, we focus on the spiny projection neurons (SPNs), the type of neuron that is primarily affected by Huntington's disease. \textbf{b.} Cell count distribution split according to condition}
    \label{fig:cell_count}
\end{figure*}

As NN applications in genomics are rapidly growing, diverse architectures of NN models are being deployed across different stages of single-cell data analysis. These include Feed Forward Neural Networks (FFNN), Auto-encoders (AE), and Transformers (a review can be found in \cite{Erfanian2023DeepAnalysis}). Nevertheless, limited research has been done exploring the applicability of XAI techniques for single-cell analysis. A notable application is presented in  \cite{zhou2024characterizing}, which employs a random forest model with SHapley additive explanations (SHAP) to identify potential indicators of preeclampsia. Additionally, \cite{yap2021verifying} trained a convolutional neural network (CNN) model to classify 47 tissue types using bulk RNA-seq data and used a SHAP gradient explainer to identify discriminatory genes. In their analysis, they found that genes identified via SHAP values were a subset of those identified using a traditional DGE analysis technique. Despite this progress, however, further exploration in this domain is needed.
Here, we propose an NN-based approach for identifying genes that contribute to a given condition of interest. Our method utilizes a NN model combined with SHAP \cite{lundberg2017unified} values to assess the contribution of individual genes to the model's predictions at single-cell resolution. By assigning importance scores to genes for each individual cell, the proposed approach provides valuable insights into the underlying biological mechanisms. 

Neurodegenerative diseases (NDD) are a group of health limiting conditions where it exist a progressive loss of brain function and general abilities. Even in cases where the inheritance of a known altered gene triggers the disease condition, the underlying mechanisms driving disease onset and progression remains unclear, which is why NDDs such as Huntington's disease (HD), lack an effective treatment. HD is a hereditary disease caused by an altered Huntingtin gene. After decades of healthy life, motor dysfunction appears in HD patients as the principal but not only HD manifestation. HD symptomatology is driven by a global loss of specific neurons, spiny projection neurons or SPNs, that populates the striatum wich is the main brain region affected in the disease. In this work, we use a snRNA-seq dataset consisting of both wild-type (WT) and HD mouse model samples \cite{lim2022huntington}. We demonstrate that training a NN classifier and evaluating feature importance with SHAP values can facilitate the discovery of altered genes that play a pivotal role in HD, which we validate with previous experimental results, and compare the approach with a traditional DGE analysis technique.
\begin{figure*}[ht!]
    \centering
    \includegraphics[width=1\textwidth]{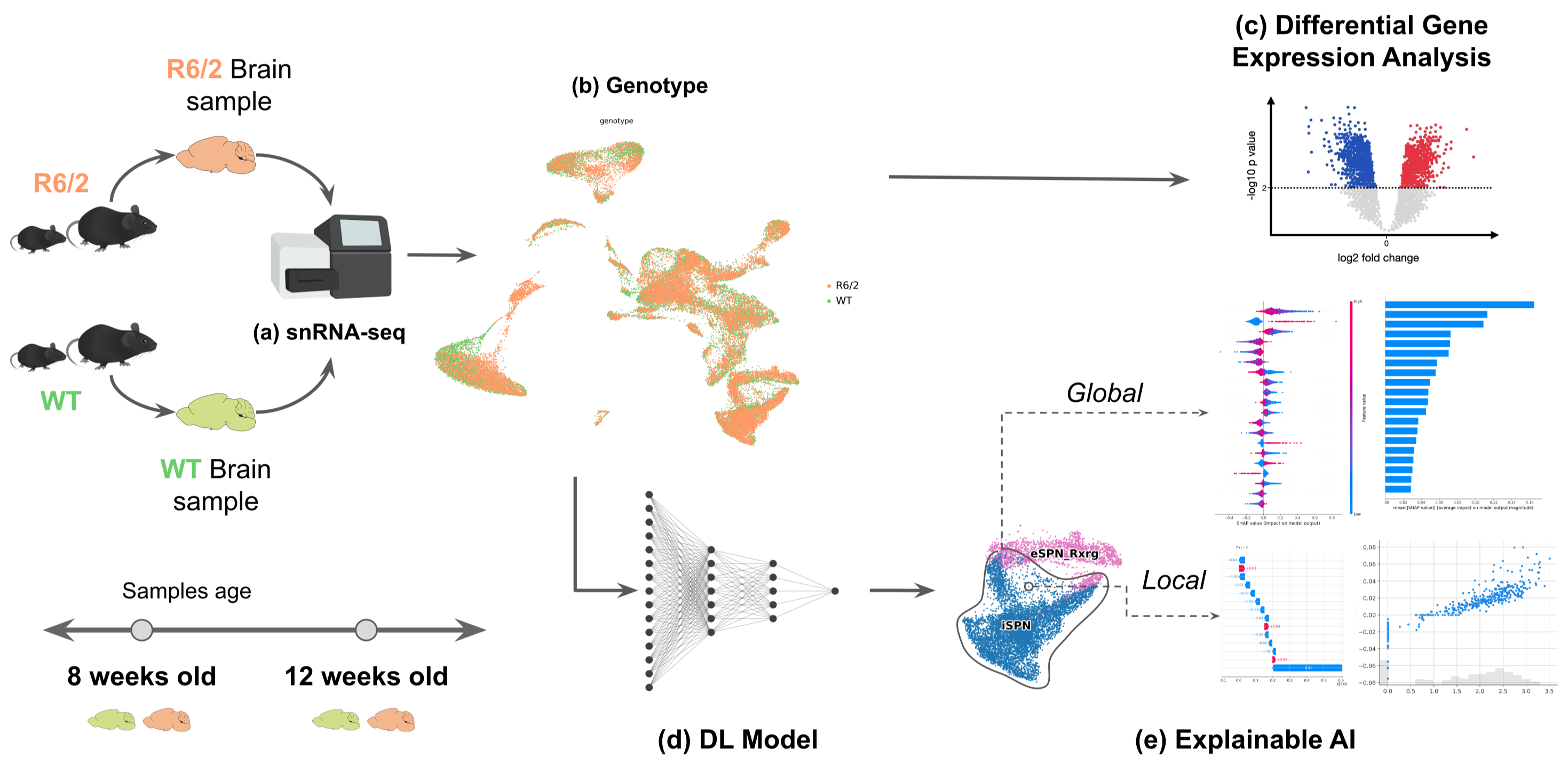}
    \caption{Model diagram of single cell analysis shows samples from R6/2 Huntington (HD) mice model and Non-transgenic (NT) which is Wild-type (WT) mice were collected at two different developmental stages 8 weeks old and 12 weeks old of the brain followed by a Single nucleus RNA Sequencing (a) is done to generate a cell atlas (b) for both conditions WT and R6/2. Subsequently, we perform differential expression analysis (c). A NN model (d) is trained on these 2 conditions combined with explainable AI (e) to identify potentially altered genes to understand disease mechanisms }
    \label{fig:model_diagram}
\end{figure*}

\section{Materials and methods}

\subsection{Dataset}

This study is based on striatal snRNA-seq obtained from two post-natal stages, 8 and 12 weeks old, from wild-type (WT) and an HD mouse model \cite{lim2022huntington}. First, we generated the single-cell gene count matrices using CellRanger \cite{cellranger}. Next, we used Seurat \cite{hao2021integrated} to normalize the counts and identify highly variable genes, resulting in a normalized matrix consisting of 42,800 cells and the top 2,500 most variable genes. Then, we clustered the cells, resulting in 17 clusters (Fig.~\ref{fig:cell_count}). Finally, we used the \verb'FindMarkers' function in Seurat to generate cluster marker genes that allowed us to assign a cell-type to each cluster based on the comparison with known specific cell-type gene markers described in the literature. This way we have identified, among the highest cell count clusters, two clusters corresponding to SPNs: cluster corresponding to indirect-pathway SPNs (iSPNs), characterized by the expression of \textit{Drd2}, \textit{Adora2a}, \textit{Penk}, and \textit{Oprd1}; and cluster corresponding to direct-pathway SPNs (dSPNs), characterized by the expression of \textit{Drd1}, \textit{Sp9}, and \textit{Sp8}. Since SPNs are the most abundant neuronal population in the striatum and specifically affected in HD \cite{albin1995selective}, we chose to focus our analysis on the aforementioned SPN clusters.

\subsection{DGE Analysis}

DGE analysis is used to analyze gene expression in cells. Mathematically, DGE seeks to estimate the density function of the expression of the $i$-th gene $Y_i$ as a function of a set of covariates $X$, i.e., $p(y_i\mid x)$. DESeq2 \cite{love2014moderated} is a commonly used tool to perform DGE which relies on a negative binomial generalized linear model. It performs a statistical test on each gene individually to identify relevant genes for a given contrast of interest (e.g., WT \textit{vs} HD), resulting in a list of effect sizes and p-values, one for each gene.  In our study, we employed DESeq2 for each of the top 2,500 highly variable genes in each cell-type independently.

\subsection{NN Classifier}

 We used a Multi-Layer Perceptron (MLP) to implement a classifier. An MLP is an FFNN typically composed of one input layer, one output layer and one or more hidden layers. In our work, we used an MLP with two hidden layers. The input layer receives the normalized gene expression values with each gene corresponding to a feature in the input vector. The hidden layers are composed of fully connected neurons and utilize the non-linear activation function ReLU (Rectified Linear Units) to model intricate relationships and interactions within the expression data. The final output layer uses a sigmoid activation function resulting in a probability of the cell being HD, which uses a threshold of 0.5 to classify cells. Thus, in contrast to traditional differential analysis approaches, this model implicitly learns the distribution $p_{\varphi}(x\mid y_1,\ldots,y_G)$ where $\varphi$ is a neural network, allowing for complex interactions and non-linearities, $y_i$ is the normalized expression of the $i$-th gene, and $x$ is the condition of interest. The initial data was split into training and test sets in an 80:20 ratio. For model training, we used 34,262 cells from various cell-types, while the test set consisted of 8,566 cells of mixed cell-types. For model evaluation, we used balanced accuracy, precision, recall and F1 score.


\begin{table*}[ht!]
\centering
\begin{tabular}{ccccccc}
 & \textbf{Precision} & \textbf{Recall} & \textbf{F1} & \textbf{WT} & \textbf{HD} & \textbf{Total Count} \\ \hline
\textbf{Cluster-wise model performance} &  &  &  &  &  &  \\ \hline
Adult NPC & 0.71 & 0.88 & 0.78 & 85 & 60 & 145 \\
Astrocyte & 0.87 & 0.88 & 0.87 & 511 & 456 & 1047 \\
Cholinergic/Adrenergic neurons & 0.83 & 0.87 & 0.85 & 565 & 482 & 1047 \\
Cortical neurons & 0.84 & 0.97 & 0.90 & 33 & 34 & 67 \\
Endothelial & 0.90 & 0.95 & 0.92 & 74 & 73 & 147 \\
Immune cells & 0.96 & 1.00 & 0.98 & 26 & 19 & 45 \\
Interneuron \textit{Nos1} & 0.83 & 0.87 & 0.85 & 23 & 26 & 49 \\
Interneuron \textit{Wdfy3} & 0.94 & 0.95 & 0.94 & 122 & 143 & 265 \\
Interneuron \textit{Whrn} & 0.87 & 0.93 & 0.89 & 169 & 185 & 354 \\
Microglia & 0.82 & 0.85 & 0.83 & 106 & 80 & 186 \\
Oligodendrocyte \textit{pik3r3} & 0.80 & 0.89 & 0.84 & 9 & 161 & 170 \\
Oligodendrocyte \textit{prr5l} & 0.90 & 0.97 & 0.93 & 811 & 604 & 1415 \\
Perivascular pericytes & 0.56 & 0.47 & 0.51 & 19 & 18 & 37 \\
dSPN & 0.98 & 1.00 & 0.99 & 527 & 738 & 1265 \\
eSPN \textit{Adam12} & 0.98 & 1.00 & 0.99 & 175 & 303 & 478 \\
eSPN \textit{Rxrg} & 0.94 & 0.98 & 0.96 & 192 & 248 & 440 \\
iSPN & 0.99 & 1.00 & 0.99 & 642 & 847 & 1489 \\ \hline
\textbf{Overall Model Performance} &  &  &  &  &  &  \\ \hline
Class HD & 0.95 & 0.91 & 0.93 & 4089 &  & 8566 \\
Class WT & 0.91 & 0.94 & 0.93 &  & 4477 & 4477
\end{tabular}
\caption{Model performance metrics for each cell-type and overall model performance, including accuracy, precision,
recall, F1 and cell counts for wild-type (WT) and Huntington’s disease (HD) cells, as well as total cell
count.}
\label{tab:model_performance}
\end{table*}

\subsection{XAI Analysis}

SHAP (SHapley Additive exPlanations) \cite{lundberg2017unified} is one of the prominent explainable AI techniques based on the game theory concept of Shapley values. In the deep learning domain, Shapley values quantify the marginal contribution of each feature to the prediction outcome, considering all possible permutations of the feature combinations as this approach ensures fairness by accounting for interactions and dependencies between features. SHAP can provide both local and global explanations of model behaviour. Mathematically, SHAP is defined as:

\begin{equation}
    \phi_i(f) = \sum_{\mathcal{S} \subseteq \mathcal{F}_i} \frac{|\mathcal{S}|!(M-|\mathcal{S}|-1)!}{M!}[f(x_{\mathcal{S}} \cup x_i) - f(x_{\mathcal{S}})]
    \label{eq:shap}
\end{equation}

where $\phi_i(f)$ represents the SHAP value for a specific feature in the model $f$, and $\mathcal{F}_i$ is the set of $M$ features excluding the $i$-th feature. Including a feature means using the actual value (gene expression) of that gene when evaluating model prediction and vice versa. The summation considers all possible subsets $\mathcal{S}\subseteq\mathcal{F}_i$, and calculates the marginal contribution of the $i$-th feature. This calculation accounts for all possible combinations of features and their interactions, providing a comprehensive understanding of how each feature influences the model's predictions. In our case, a positive SHAP value implies that the expression of the corresponding gene contributes to HD phenotype, whereas a negative value contributes to WT phenotype. 

To conduct the XAI analysis, we used the KernelExplainer from SHAP, which uses a special weighted linear regression to compute the importance of each feature. An explainer was created using the training data set as the background to generate explanations for the HD cells in each cluster from the test set. We used this approach to identify the set of \textit{informative genes} driving the prediction for this genotype. Furthermore, we took advantage of the single-cell resolution to compute the Pearson correlation coefficient between the gene expression and the individual SHAP values.

\section{Results}

\subsection{Classifier performance}

We evaluated the performance of the NN classifier in distinguishing between the two classes across the entire dataset on the test set looking at overall and cell type-specific metrics (Table \ref{tab:model_performance}). The NN model demonstrates high F1 score across most clusters, with the highest F1 score of 0.99 observed in the iSPN cluster and 0.992 in dSPN cluster. This indicates that the model performs exceptionally well in distinguishing between WT and HD cells in SPNs, particularly affected in HD. 
In contrast, the NN model shows low classification performance for Perivascular pericytes. We hypothesize this could be due to the fact that there are no alterations for this cell-type in HD.
Overall, the NN model's performance evaluated just on HD cells achieves a precision of 0.95 and recall of 0.91, resulting in an F1-score of 0.93. This indicates that the model is highly effective at identifying HD cells with high F1 and low false positive rates. Similarly, when evaluating the model just with WT cells, the precision is 0.91, with a recall of 0.94 and an F1-score of 0.93. These metrics suggest a well-balanced performance in identifying WT cells, with a slightly higher recall compared to precision. With an overall accuracy of 0.93, the model shows a robust performance.

\begin{figure*}[ht]
    \centering
    \begin{subfigure}[b]{0.45\textwidth}
        \centering
        \includegraphics[width=\textwidth]{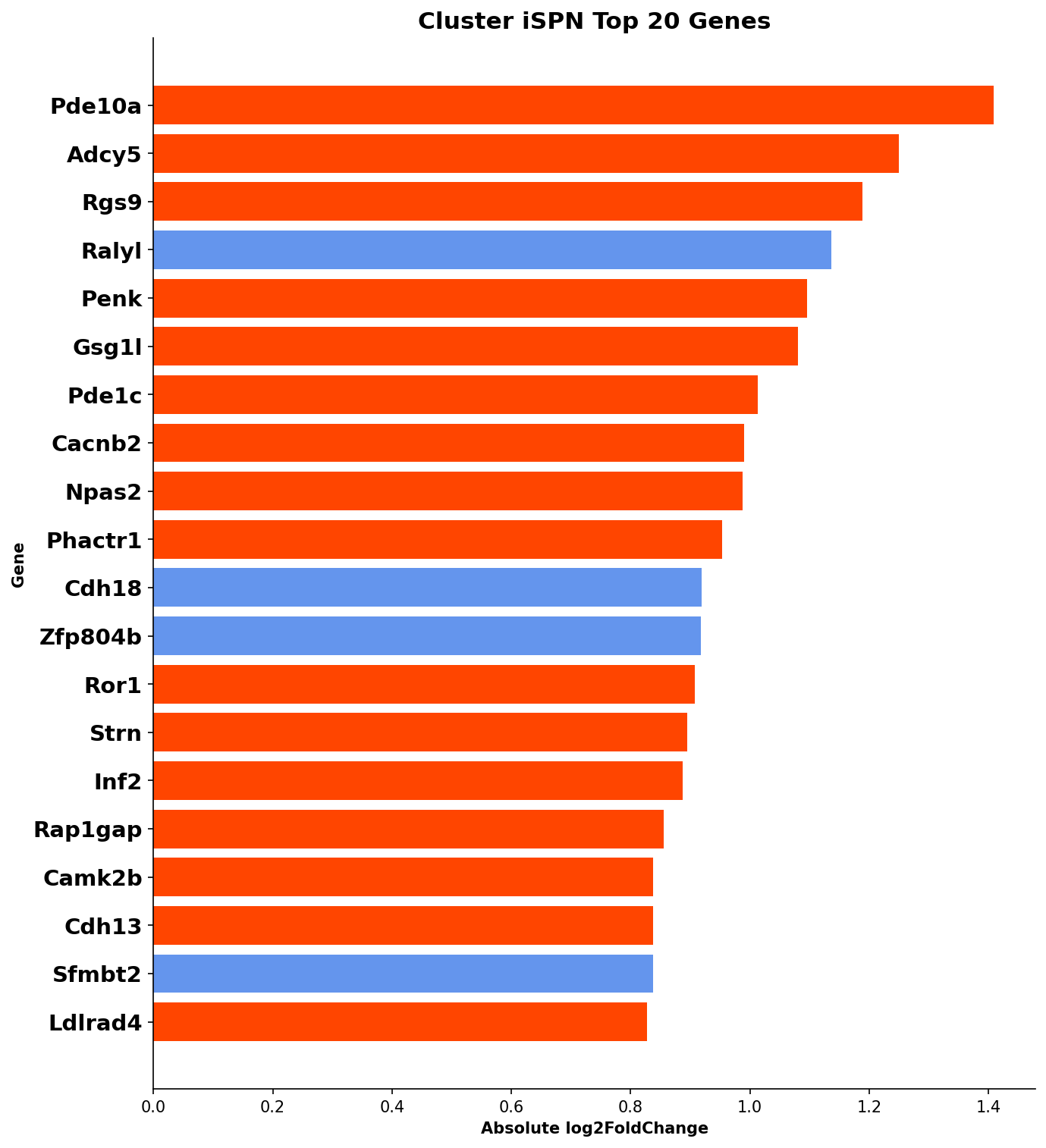}
    \end{subfigure}
    \hfill
    \begin{subfigure}[b]{0.45\textwidth}
        \centering
        \includegraphics[width=\textwidth]{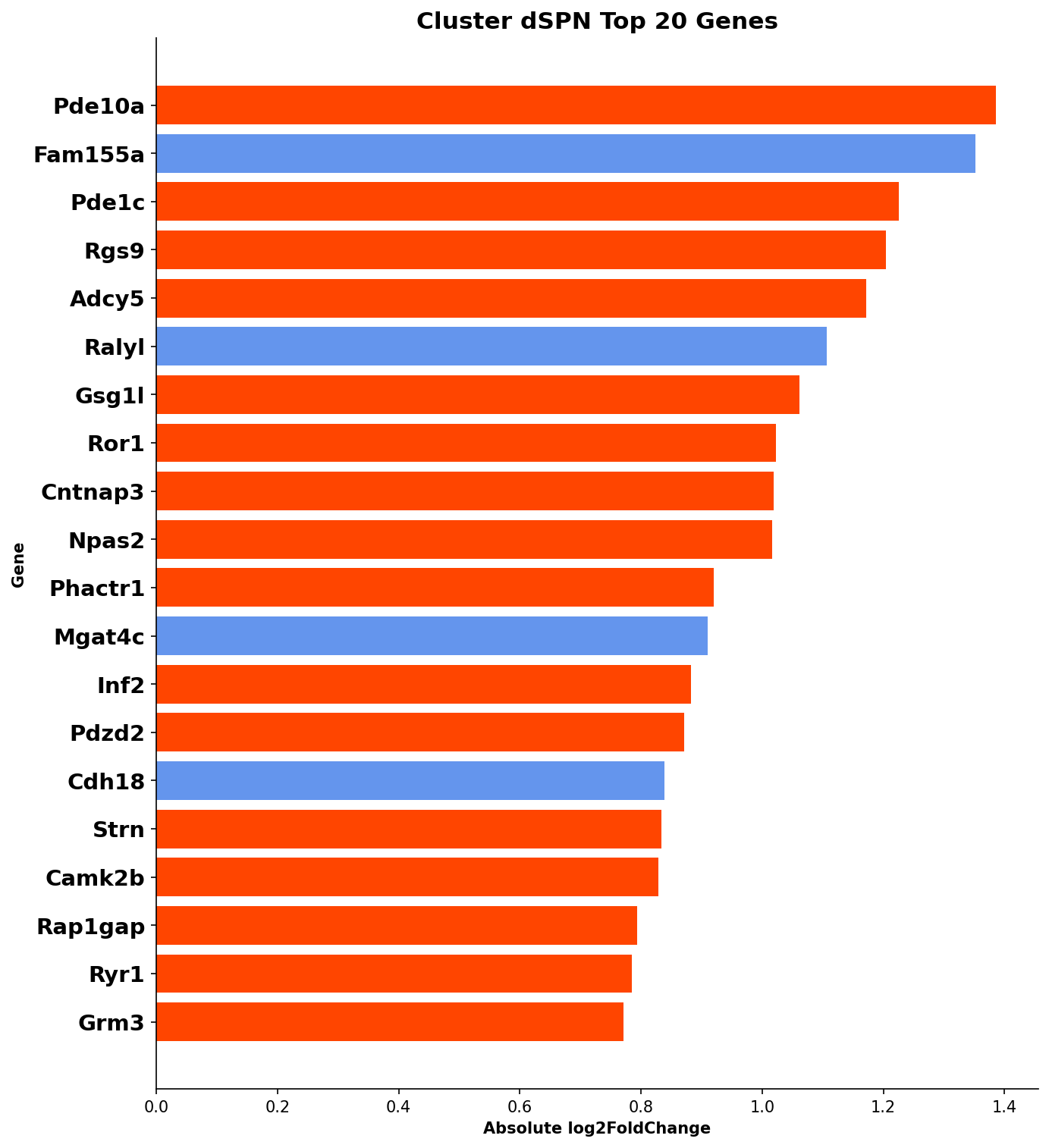}
    \end{subfigure}
    \caption{Barplot displaying top 20 DEGs from DESEq2 based on absolute LFC for clusters iSPN (left) and dSPN (right). Bars are colour-coded to indicate HD upregulated genes (blue) and down-regulated (red).}
    \label{fig:deseq2_top20}
\end{figure*}

\begin{figure*}[t]
    \centering
    \begin{subfigure}[b]{0.45\textwidth}
        \centering
        \includegraphics[width=\textwidth]{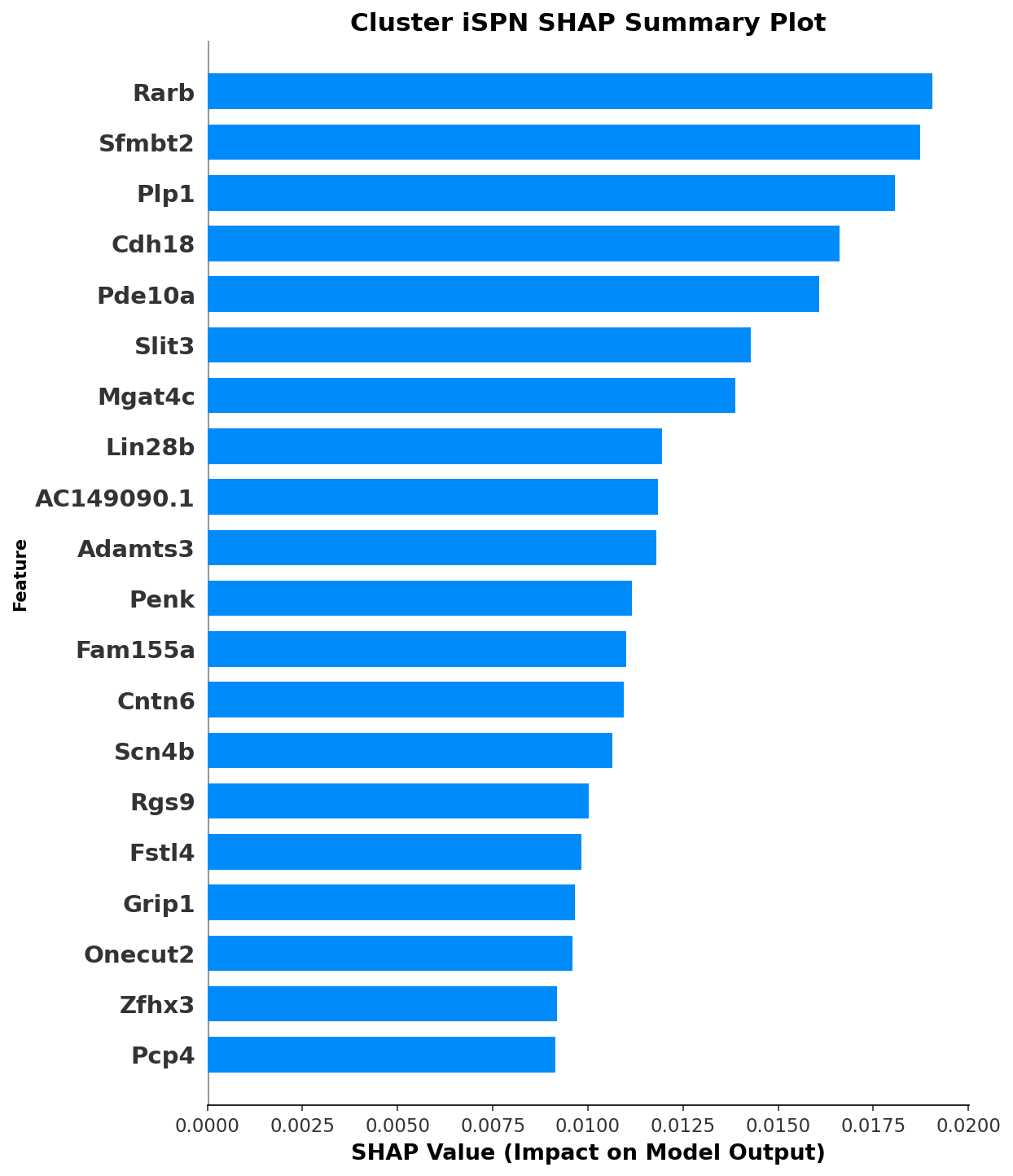}
    \end{subfigure}
    \hfill
    \begin{subfigure}[b]{0.45\textwidth}
        \centering
        \includegraphics[width=\textwidth]{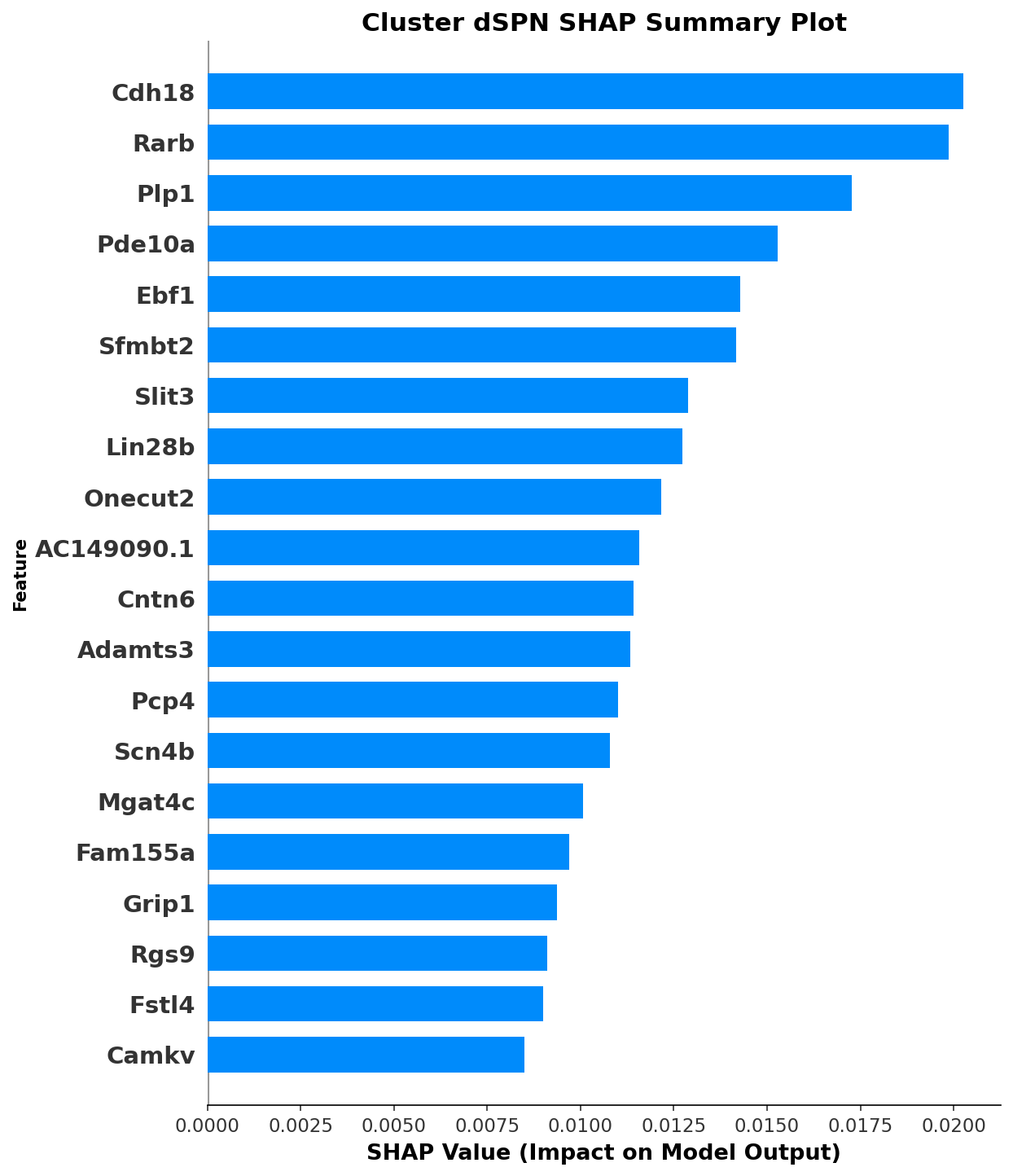}
    \end{subfigure}
    \caption{SHAP summary plot showing gene importance ordered by mean absolute SHAP values for HD cells in clusters iSPN (left) and dSPN (right).}
    \label{fig:shap_summary_bar}
\end{figure*}

\subsection{Top genes identified by DESeq2 and SHAP}

To evaluate the performance capabilities of DESeq2 and SHAP methods in identifying relevant HD-related genes first we compared the top 20 genes obtained for clusters iSPN and dSPN using each method (Fig.~\ref{fig:deseq2_top20} and \ref{fig:shap_summary_bar}).

\begin{figure*}[ht]
    \centering
    \begin{subfigure}[b]{0.30\textwidth}
        \centering
        \includegraphics[width=\textwidth]{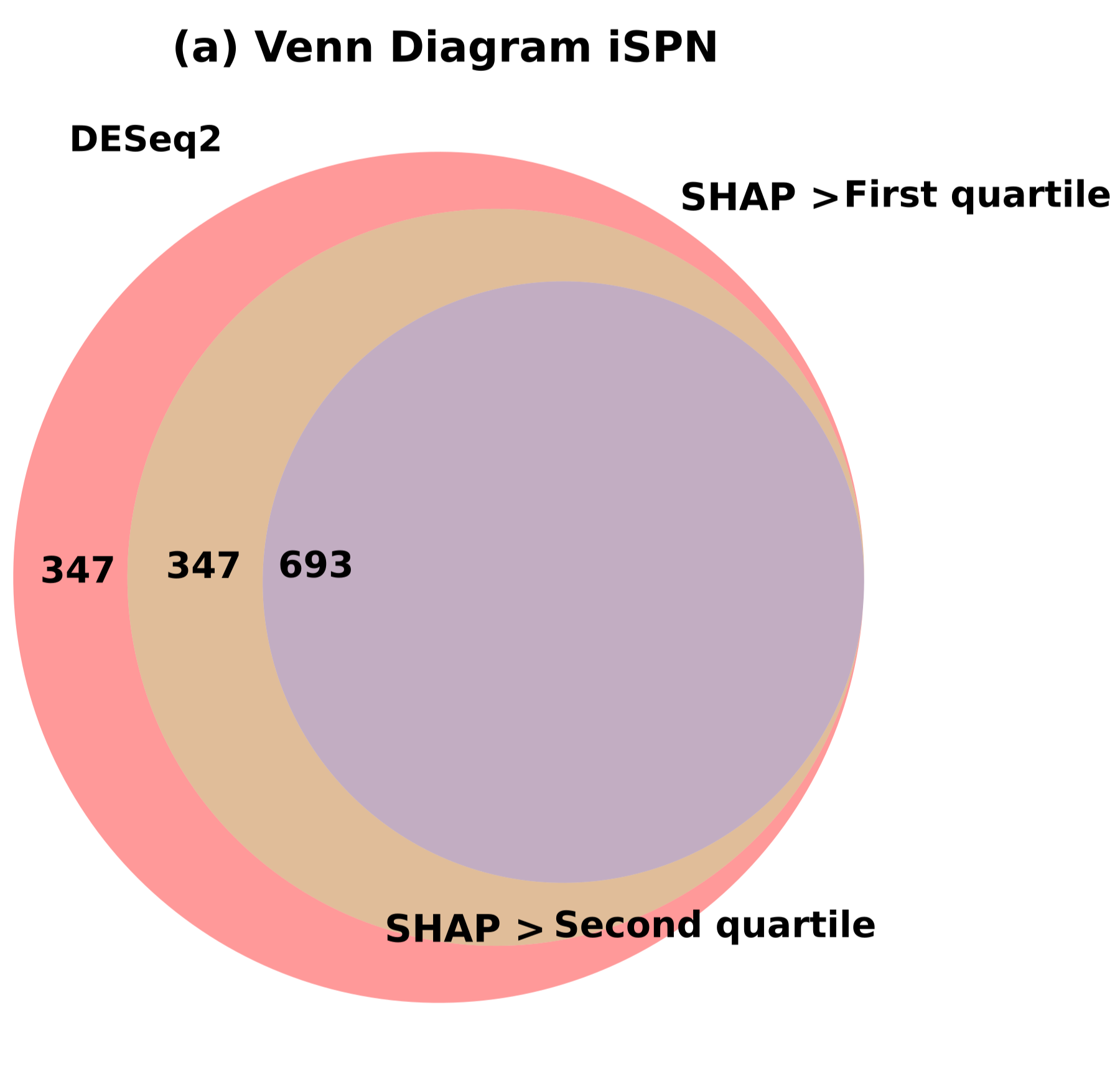}
    \end{subfigure}
    \hspace{0.12\textwidth} 
    \begin{subfigure}[b]{0.30\textwidth}
        \centering
        \includegraphics[width=\textwidth]{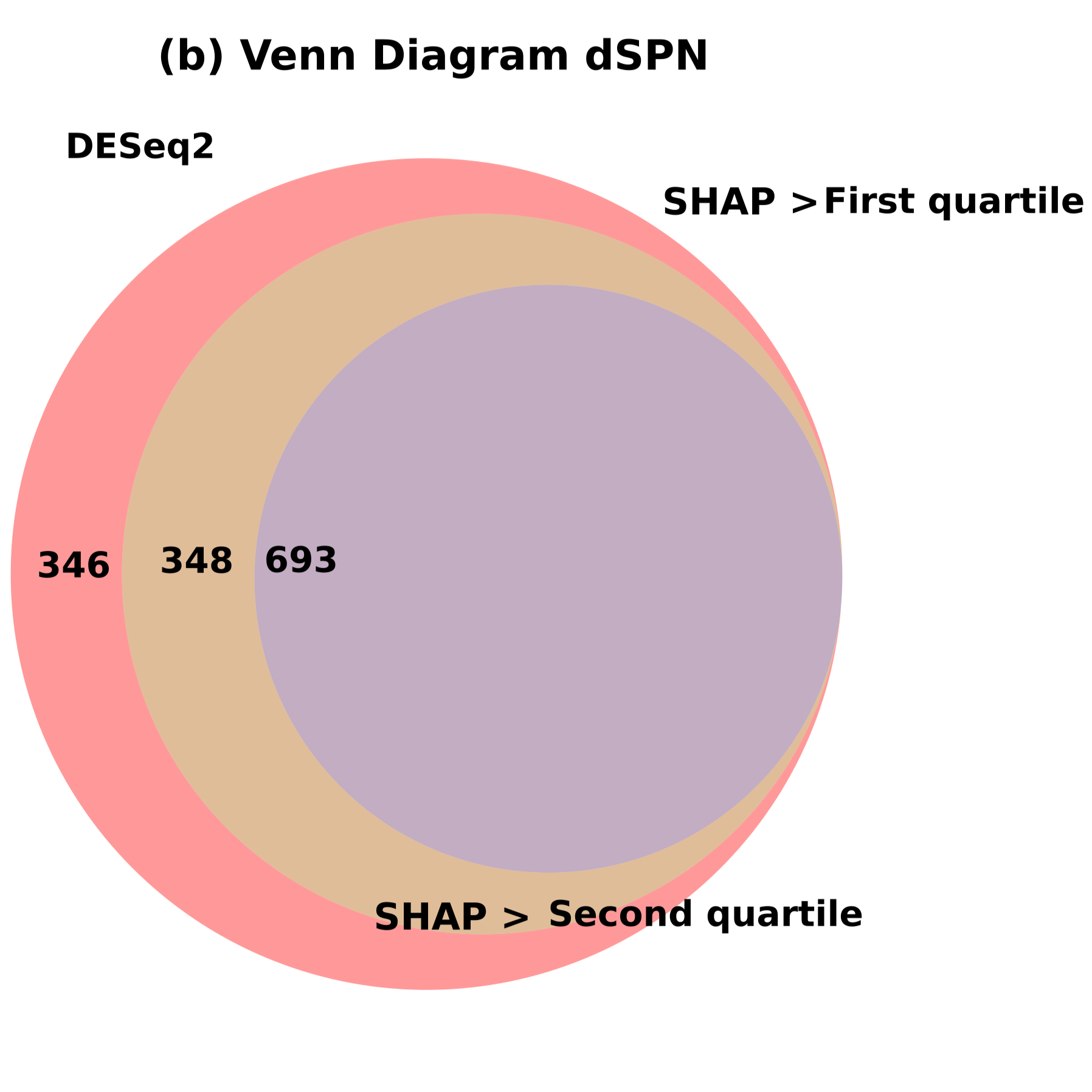}
    \end{subfigure}
    \caption{Venn diagram illustrating the overlap between differentially expressed genes identified by DESeq2 and informative genes identified by SHAP values. The diagrams show the intersection of these gene sets when applying thresholds on the mean absolute SHAP values that correspond to it's quartiles for (a) cluster iSPN, with SHAP value $>$ first quartile and SHAP value $>$ second quartile, and (b) cluster dSPN, with SHAP value $>$ first quartile  and SHAP value $>$ second quartile.}
    \label{fig:Venn_diagram}
\end{figure*}

To compare the ranking of genes obtained by both methods, we calculated the Spearman correlation. For the iSPN cluster, the Spearman correlation between mean SHAP values and DESeq2 log fold-change (LFC) was found to be 0.286 ($p$-value = 1.20e-27). Similarly for cluster dSPN, the Spearman correlation was 0.283 ($p$-value = 4.05e-27). These results suggest that there is only a partial agreement in the gene rankings generated by both methods. To visualize how genes identified using DESeq2 intersect with that of SHAP, we generated Venn diagrams using various thresholds based on quartiles of mean absolute SHAP values. The results are shown in Fig. \ref{fig:Venn_diagram}.

Entering into gene details, both methods were able to identify common genes for each cluster studied (\textit{Pde10a}, \textit{Cdh18}, \textit{Penk} \textit{Sfmbt2} for cluster iSPN; \textit{Pde10a}, \textit{Cdh18}, \textit{Fam155a},  \textit{Rgs9} for cluster dSPN). From these genes, we identify new HD-related genes that are common to both methods: \textit{Cdh18} on neuronal SPNs clusters (SAHP: iSPN, dSPN, eSPN, adult NPC), \textit{Fam155} encoding a component of the sodium selective NALCN multi-protein complex \cite{kang2020structure} that is found to be generally altered in different cell-types (SPNs, AdultNPC, Oligondendrocytes, Interneuron, Astrocytes) showing both a positive SHAP correlation.
Interestingly, however, we find genes that are only identified when using the proposed SHAP analysis. Some of these have been previously described in the context of HD, including \textit{Rarb} \cite{zinter2024compromised}, \textit{Cntn6} genes\cite{christodoulou2020investigating} or \textit{Onecut2} \cite{le2017altered}. Importantly, we also identified informative genes that have not been previously described in the context of HD, such as \textit{Slit3}, showing a SHAP positive correlation in both SPN clusters.

Further model validation was made from SHAP results based on published data obtained from \textit{in vivo} HD models or patients. For this validation, we compare the correlation between SHAP values and gene expression of previously described altered genes, expecting to observe a positive correlation between higher HD expression levels with positive SHAP values and a negative correlation between reduced HD expression levels with negative SHAP values. Among the top 20 informative genes, some were previously described as being upregulated in HD. For them, we observe a positive correlation between SHAP and gene expression in SPN clusters are: \textit{Onecut2} \cite{le2017altered} or \textit{Sfmtb2} \cite{becanovic2010transcriptional}. Genes that have been previously shown to be downregulated in HD from which we obtain a negative correlation between SHAP and gene expression in SPN clusters are: \textit{Pde10A} \cite{Niccolini2015} or \textit{Scnb4} \cite{Oyama2006https://doi.org/10.1111/j.1471-4159.2006.03893.x}. We have validated other lower rank genes previously described in different HD models, such as \textit{Penk1} (cluster iSPN), \textit{Gria3} (clusters iSPN, dSPN) \cite{Hodges2008} or \textit{Nrg1} \cite{siebzehnrubl2018early}. Although our main focus was put on SPN clusters, other neuronal clusters are worth mentioning since they show shared altered genes with SPN clusters, eg. \textit{Gria3} and \textit{Nrg1} in clusters eSPN and adult NPC,\textit{ Onecut2} in eSPN \textit{Adam12}, \textit{Pcp4} in eSPN \textit{Rxrg} and Adult NPC, or \textit{Pde10A} in eSPN, Interneuron \textit{Whrn}, Cholinergic Adrenergic neurons.

Notably, SHAP allows a deeper interpretation on gene expression level compared to DESeq2. For instance, \textit{Fam155} shows a different contribution to HD probability depending on the level of expression, being low \textit{Fam155} levels negatively correlated to HD probability, whereas higher levels are positively correlated, therefore suggesting the requirement of a fine control at the cellular level of its expression to increase WT probability correlation.


\begin{table*}[ht!]
\centering
\begin{tabular}{llll}
\textbf{SHAP} & \textbf{NES} & \textbf{DESeq2} & \textbf{NES} \\ \hline
\multicolumn{4}{c}{\textbf{Cluster iSPN}} \\ \hline
MIR 7119 3P & 3.31 & MIR 7A 1 3P & 2.47 \\
MIR 218 5P & 3.05 & MIR 677 5P & 2.39 \\
MIR 7661 5P & 2.52 & MCCLUNG CREB1 TARGETS DN & -2.22 \\
MIR 300 3P & 2.44 & GOBP MONOATOMIC CATION TRANSPORT & -2.33 \\
LEIN MIDBRAIN MARKERS & 2.31 & EED TARGET GENES & -2.87 \\
GOBP ENSHEATHMENT OF NEURONS & 2.30 & GOCC SYNAPSE & -3.99 \\
GOCC NEURON TO NEURON SYNAPSE & -2.39 &  &  \\ \hline
\multicolumn{4}{c}{\textbf{Cluster dSPN}} \\ \hline
MIR 218 5P & 3.41 & GOMF RNA BINDING & 2.64 \\
MIR 7119 3P & 2.95 & MIR 677 5P & 2.47 \\
LEIN MIDBRAIN MARKERS & 2.60 & ZFP319 TARGET GENES & 2.45 \\
MIR 1907 & 2.48 & GOCC SYNAPSE & -3.26 \\
GOMF PHOSPHORIC DIESTER HYDROLASE ACTIVITY & -2.30 &  & 
\end{tabular}
\caption{Gene Set Enrichment Analysis (GSEA) using a ranked list of genes from SHAP and DESeq2 with Normalized enrichment score (NES) for clusters iSPN and dSPN}
\label{tab:GSEA_analysis}
\end{table*}
\subsection{GSEA results for SHAP and DESeq2}
Next, Gene Set Enrichment Analysis (GSEA) was carried out in the SPN clusters (dSPN, iSPN) to gain deeper insights into the biological significance of the genes identified by both SHAP and DESeq2. GSEA helps us understand the dysregulated pathways in each cell-type and provides a comprehensive view of the underlying HD biological processes.

GSEA results from both DESeq2 and SHAP are shown in Table \ref{tab:GSEA_analysis}. Common to both methods, we find a decreased synaptic function in SPN clusters. However, GSEA from DESeq2 data emphasizes the decrease in a global broader list of synaptic-related regulators included in the category named \textit{Synapse}, whereas GSEA from SHAP manifest a decrease specifically in \textit{Neuron-to-Neuron Synapse}, remarking regulatory pre and postsynaptic proteins important for synaptic function and plasticity, which has been previously described as HD relevant in \cite{Smith-Dijakhttps://doi.org/10.1111/jnc.14723}. In both cases, in this synapse-related category well-known HD genes are included, such as \textit{Drd1} (marker for dPSN), \textit{Drd2} or \textit{Adora2a} (both markers for iSPN).

Following the initial analysis, we identified downregulated and upregulated pathways by GSEA from DESeq2 data. More specifically, in cluster iSPN we observe the downregulation of three gene categories directly related to neuronal function and SPN identity: first, \textit{Monotatomic Cation Transport} category mainly consisting in ion transporters such as calcium, potassium and synaptosome proteins (Cacna, Kcn, Scn4b or Vamp family proteins), respectively previously shown to be decreased in a HD model \cite{SAPP2020104950}); second, \textit{CREB1 targets}, among which known decreased SPN markers are found such as \textit{Six3}, \textit{FoxP1}, \textit{Penk}; and third,  \textit{EED target genes} where we found known HD/striatal genes such as \textit{Pax6}, \textit{Drd1} or \textit{FoxP1}.

Upregulated signalling from GSEA applied to DESeq2 includes: i) the \textit{Zfp319 target genes} category in dSPNs among which \textit{Drd1}, \textit{Ebf1}, \textit{Opcml}, \textit{Meis2} are listed, all these being known contributors to SPN identity and function; ii) \textit{RNA binding proteins}, indicating HD abnormally increased protein translation, as previously described in \cite{creus2019increased}; and iii) several \textit{MIR-predicted target} genes (MIR-677-5p) in both dSPN and iSPN clusters, or MIR-7a-1-3p in cluster iSPN) whose implication in HD are not described to date. Remarkably, MIR-677-5p is differentially expressed in an AD mouse model \cite{song2021microarray}.

We next focused on the outcome obtained exclusively using SHAP data. We find here the HD-downregulation of \textit{Phosphoric diester hydrolase activity} category on cluster dSPN. This category encompasses known HD-related genes such as \textit{Pde10a}, a protein that coordinates striatal signaling and has been proposed as HD early biomarker \cite{Niccolini2015} or \textit{Fan1} \cite{mcallister2022exome}, a DNA repair enzyme whose variants altere HD disease progression.   
Other HD-upregulated pathways only found with SHAP are: i) \textit{Midbrain markers} at both SPN clusters, whose HD implication need to be further explored, ii) several \textit{MIR target genes} categories can also be found from GSEA applied on SHAP obtained data, e.g. MIR 218-5p, MIR 1907 for cluster dSPN, and MIR218-5p, MIR7661-5p, MIR300-3p for cluster iSPN, or MIR 7119-3p for both SPN clusters. To our knowledge, they have not been associated to HD, thus further exploration of those MIRs or their targets is required. Interestingly, MIR 218-5p is an epigenetic modifier whose expression levels are associated with an increased in stress susceptibility and impact on synaptic function and plasticity \cite{schell2022mir}. Another disease-related MIR detected using SHAP is MIR 300-3p which has been previously seen to be upregulated after ischemic stroke and proposed as a biomarker of this condition \cite{li2018plasma}.  
One interesting HD-upregulated category only found in cluster iSPN when analyzing SHAP results is the so call \textit{Ensheathment of neurons}. Here, we can find genes encoding adhesion molecules and myelin related factors such as \textit{Ptprz1} with a regulatory role on inflammation and remyelination processes in the brain \cite{fujikawa2016role, nagai2022protein}.
In summary, GSEA manifests different HD relevant signalling pathways depending on the initial model used, remarking the relevance of complementing traditional models with XAI models to have a complete view of the mechanisms involved in a particular disease. In particular, GSEA applied to DESeq2 data reflects the global synaptic function and cation transport decrease and the upregulation of both RNA binding proteins and MIRs associated with neurodegeneration. On the other hand, GSEA applied to SHAP analysis shows up the upregulation of relevant stress-related MIRs and downregulation of synaptic function and plasticity related genes.

 \section{Discussion}

Commonly used DGE techniques such as DESeq2 provide useful insights into gene expression changes between conditions. However, these techniques are not able to capture the association and interaction of genes at different levels. In contrast, the XAI-based approach adopted in this paper is able to analyze how individual genes contribute to the overall disease state, both locally at single cell level or globally at cell-type level, being this ability one of the major strengths of the method. When analyzing the results, there were differences in the relevant genes and pathways identified by both techniques. In spite of the fact that many genes considered relevant by DESeq2 were also identified by SHAP (see Fig. \ref{fig:Venn_diagram}), their ranking was different, as demonstrated by the relatively low value of the Spearman correlation. This would be particularly relevant to explain why both methods did not identify the same pathways, due to the fact that GSEA relies on gene rankings to generate its results.

Among SHAP-based identification of relevant HD genes, we find gene alterations supported by the literature. This validation opens up the possibility to propose new HD contributing genes and altered mechanisms to be further explored with \textit{in vivo} experiments. Since the altered pathways highlighted by each approach using GSEA vary depending on the model, new perspectives of disease-related mechanisms can be found with the proposed combination of NN model with SHAP. 

\section{Conclusions}

The use of NN models with XAI techniques offers a more detailed analysis of gene expression at single-cell resolution when compared with traditional techniques. A subset of genes and altered pathways are only detected using the proposed XAI approach and are missed by a traditional differential expression method, which underestimates their potential contribution to the disease. Future research should be focused on further investigation and characterization of the differences in the results produced by traditional DGE and XAI techniques and their biological relevance, being the XAI techniques not limited to SHAP. Other interesting directions would include, introducing more data modalities, adding more features into the training data like RNA velocity and use of more sophisticated NN models. 
In this work, we have first analyzed postnatal HD data from a mouse model, then validated the model,
and finally proposed new relevant HD genes.
Insights gained from this study can be extended to other diseases whose mechanisms are yet to be clarified. 

\section{Acknowledgment}
This study was supported by grants from the Ministerio de Ciencia e Innovación (PID2021-126961OB-I00, PLEC2022-009401); Instituto de Salud Carlos III, Ministerio de Ciencia e Innovación and European Regional Development Fund (ERDF A way of making Europe) (Red de Terapias Avanzadas, RD21/0017/0020); European Union NextGeneration EU/PRTR; Generalitat de Catalunya (2021 SGR 01094); ``la Caixa'' Foundation under the grant agreements LCF/PR/HR21-00622 and LCF/BQ/PI24/12040007; and Red Española de Supercomputación (RES) under project BCV-2024-2-0010.

\bibliographystyle{ieeetr}
\bibliography{ref}

\end{document}